\documentstyle[preprint,aps,prl]{revtex}
\begin{document}
\bibliographystyle{prsty}
\title{\Large\bf Local moment formation in quantum point contacts}
\author{Kenji Hirose$^{1}$, Yigal Meir$^{2,3,\dagger}$, 
and Ned S. Wingreen$^{2}$}
\address{$^{1}$ Fundamental Research Laboratories, NEC Corporation, 34
Miyukigaoka, Tsukuba, Ibaraki 305-8501, Japan\\
$^{2}$ NEC Research Institute, 4 Independence Way, Princeton,
New Jersey 08540\\
$^{3}$ Physics Department, Princeton University, Princeton, New Jersey
08540\\}
\date{\today}
\maketitle
\begin{abstract}
{Spin-density-functional theory of quantum point contacts (QPCs)
reveals the formation of a local moment
with a net of one electron spin in the vicinity of the point contact --
supporting the recent report of a Kondo effect in a QPC.
The hybridization of the local moment to the leads decreases as
the QPC becomes longer, while the onsite Coulomb-interaction energy
remains almost constant. }
\end{abstract}

The discovery of the extra conductance plateaus at $0.7(2e^2/h)$ in
quantum point contacts (QPCs) \cite{Thomas,Kristensen,Nuttinck} and at
$0.5(2e^2/h)$ in clean
quantum wires\cite{Reilly}, in addition to the usual quantization of the
conductance into steps of $2e^2/h$\cite{vanWees,Wharam}, has focused
attention on the role of electron-electron interaction in these
low-dimensional quantum
systems\cite{Wang98,Flambaum,Spivak,Rejec,Bruus,Tokura}.

A recent experiment on QPCs has revealed several features
characteristic of the Kondo effect\cite{Cronenwett}.
Specifically, there is a zero-bias
peak in the differential conductance which splits in an in-plane
magnetic field\cite{inplaneB}.
A single energy scale, the ``Kondo" temperature $T_K$, sets the
width of the zero-bias peak, the magnetic field required to split
the peak,  and the crossover temperature to perfect transmission.
Moreover, the extra conductance plateau at $0.7 (2e^2/h)$ can be
explained
using an Anderson model in which the hybridization of a localized
electron to the band depends on energy and valence\cite{Meir}.
A puzzling question is how a localized spin can form in an open
QPC system.

In this work, we employ spin-density-functional theory (SDFT), and
 find that a local moment
with a net of one electron spin is formed in the vicinity of a QPC.
The splitting between the two spin directions
at high in-plane magnetic fields\cite{inplaneB} approaches a finite
residual
splitting at zero field, which we interpret as the Coulomb-interaction
energy $U$ between electrons at the site of the QPC.
>From the width of the localized state we infer that the
hybridization to the leads decreases for longer QPCs, while
the Coulomb $U$ remains nearly constant.
Our interpretation that a {\it dynamical} local moment forms
at the QPC differs from that of Wang and Berggren\cite{Wang98}
who infer a true spin-polarized ground state. We believe that the
experimental observation of signatures of the Kondo
effect\cite{Cronenwett}, combined with a theorem
forbidding spin-polarization in one dimension\cite{Lieb}, strongly
favor the dynamical-spin interpretation.

The electronic states of a realistic quantum point contact
are obtained using
density-functional theory within the local-density
approximation\cite{Callaway}. This method allows us to study both
potential confinement and electron-electron interaction in a unified
framework.
First, for a clean quantum wire uniform in the $x$-direction
and with a parabolic confining potential in the $y$-direction of
$V_{\rm ext}^{0}(y)=(1/2)m^{*}\omega_y^2 y^2$, we readily
find the wave functions
$e^{\pm ik^{\sigma}_{n} x}\psi^{\sigma}_{n}(y)$, chemical potential
$\mu$, band-edge energies $\epsilon^{\sigma}_{n}$, and the charge
density $\rho^{0\sigma}(y)$ for a given  density
$n_{1D}$
\cite{Hirose}.
Here $n$ is the subband index and $\sigma$ is spin index, respectively.

Next, we introduce the QPC potential,
\begin{equation}
V_{\rm QPC}(x,y)=V(x)/2 +
m^{*}\left(\omega_y+V(x)/\hbar\right)^2 y^2/2 - V_{\rm ext}^{0}(y)
\end{equation}
(see insets to Fig.~1),
where $V(x) = V_0/\!\cosh^2(x/d)$, with decay length
$d=\sqrt{2V_0/m^{*}}/\omega_x$,
 and solve the following
Kohn-Sham equation\cite{Kohn} for wave functions with energy $\epsilon$,
\begin{eqnarray}
&&\left[-\frac{\hbar^2}{2m^*}\frac{\partial^2}{\partial x^2}
+V_{\rm QPC}(x,y)+\delta V_{H}(x,y)+\delta
V^{\sigma}_{xc}(x,y)\right.
\nonumber \\
&&+\left.\frac{}{}g\mu_{B}B_{||}\sigma+\epsilon^{\sigma}_{n} \right]
\Psi^{\sigma}_{n,k_n}(x,y)=\epsilon \Psi^{\sigma}_{n,k_n}(x,y).
\label{KS}
\end{eqnarray}
Electrons incident from the two leads ($x \rightarrow \pm \infty$)
are scattered elastically by the effective QPC potential $\delta
V^{\sigma}(x,y)=V_{\rm QPC}(x,y)+\delta V_{H}(x,y)
+\delta V^{\sigma}_{xc}(x,y)$, which is the difference between the QPC
self-consistent potential and that of the clean wire. The Hartree and
exchange-correlation parts of the potentials are written, respectively,
as
\begin{eqnarray}
&&\delta V_{H}(x,y)=\frac{e^2}{\kappa}\int\!\int dx'dy'
\left\{\rho(x',y')-\rho^{0}(y')\right\} \nonumber \\
&\times& \left[\frac{1}{\sqrt{(x-x')^2+(y-y')^2    }}
            -\frac{1}{\sqrt{(x-x')^2+(y-y')^2+a^2}}\right], \\
&&\delta V^{\sigma}_{xc}(x,y)=
    \frac{\delta E_{\rm xc}[\rho  ,\xi  ]}{\delta\rho^{ \sigma}({\bf
r})}
   -\frac{\delta E_{\rm xc}[\rho^0,\xi^0]}{\delta\rho^{0\sigma}(   y
)},
\end{eqnarray}
where we have introduced an image-charge plane at a distance $a=100{\rm
nm}$ to model the experimental geometry\cite{Hirose}. The
exchange-correlation energy functional $E_{\rm xc}[\rho,\xi]$ is treated
in the local density approximation
with the local spin polarization
$\xi({\bf r})=[\rho^{\uparrow}({\bf r})-\rho^{\downarrow}({\bf r})]
/\rho({\bf r})$\cite{local}.

The eigenstates of (\ref{KS}) can be characterized as waves incident
from the left $\Psi_{n,k^{\sigma}_{n}}^{L}(x,y)$ and
from the right $\Psi_{n,k^{\sigma}_{n}}^{R}(x,y)$.
Expanding,
$\Psi_{n,k^{\sigma}_{n}}^{L}(x,y)=\sum_{m}u_{n,m,k^{\sigma}_{n}}^{L}(x)
\psi^{\sigma}_{m}(y)$, $u_{n,m,k^{\sigma}_{n}}^{L}(x)$ has the form of
plane waves with wavevectors
$k^{\sigma}_{n}=\sqrt{2m^{*}(\epsilon-\epsilon^{\sigma}_{n})}/\hbar$
far from the QPC region:
\begin{equation}
u_{n,m,k^{\sigma}_{n}}^{L}(x)=\left\{
\begin{array}{cll}
e^{i k^{\sigma}_{n} x}\delta_{n,m}+r^{\sigma}_{n,m}e^{-i k^{\sigma}_{m}
x};
& \quad x \leq -x_0 & \quad \\
t^{\sigma}_{n,m}e^{i k^{\sigma}_{m} x}; & \quad x \geq x_0
\end{array}
\right.
\label{bound1}
\end{equation}
where $r^{\sigma}_{n,m}$ and $t^{\sigma}_{n,m}$ are the elements of
unknown reflection and transmission matrices, and where $|x_0| = 500
{\rm nm}$ is sufficiently
far from the QPC that $\delta V(x,y)$ is negligible.
$\Psi_{n,k^{\sigma}_{n}}^{R}(x,y)$ is expanded analogously.
We employ the recursion-transfer-matrix (RTM) method\cite{Hirose2} to
solve the Kohn-Sham equation (\ref{KS}) with the above boundary
conditions.
>From the resulting wave functions, the density $\rho(x,y)$ is
constructed
as a sum over occupied states,
\begin{equation}
\rho(x,y)=\frac{1}{2\pi}\sum_{{n,\sigma}\atop{L/R}} \int_{0}^{\infty}\!
f\left(\epsilon^{\sigma}_{n}+\frac{\hbar^2 k^{\sigma
2}_{n}}{2m^*}\right)
|\Psi^{L/R}_{n,k^{\sigma}_{n}}(x,y)|^2 dk^{\sigma}_{n},
\end{equation}
where the Fermi distribution function is $f(\epsilon)=1/[{\rm exp}
\left\{(\epsilon-\mu)/k_B T\right\}+1]$.
These procedures are iterated until self-consistent solutions are
obtained
for the electron density $n_{1D}$.
We use the material constants for GaAs, $m^*=0.067m_0$, $\kappa=12.9$,
$g=0.44$ and fix the external confinement in the y-direction in the wire
such as
$\hbar\omega_y=2.0{\rm meV}$.

Figure 1 shows the electronic properties of QPCs with $V_0 = 3.0{\rm
meV}$
at $T=0.1K$ for three different lengths:
(a),(d) $d=82.6$nm ($\hbar\omega_x$=1.0meV),
(b),(e) $d=55.0$nm ($\hbar\omega_x$=1.5meV), and
(c),(f) $d=41.3$nm ($\hbar\omega_x$=2.0meV).
The density far from the QPC is taken as $n_{1D}=2.80\times 10^{-2}{\rm
nm}^{-1}$,
where the electrons far into the wire are unpolarized and only the
lowest
two spin subbands contribute to transport\cite{Hirose}. We find
that a solution
with broken spin-symmetry coexists with an unpolarized solution.
To obtain the broken-symmetry solution, we first apply an
in-plane magnetic field
of up to $B=6T$\cite{inplaneB}, and then solve the Kohn-Sham equations
self-consistently while reducing the magnetic field to zero.

In Fig.~1(a)-(c) we show the self-consistent QPC barrier as a function
of
position $x$ in the direction of current flow. Specifically, we plot
the energy of the bottom of the lowest 1D subband
$\epsilon^{\sigma}_0(x)$ relative to the band-edge
$\epsilon^{\sigma}_0$  far into the wire, for both spin-up (solid lines)
and spin-down (dashed lines) electrons.  Small Friedel
oscillations with a period of $2\pi/2k_f\simeq 72{\rm nm}$
are present far into the wire. The self-consistent QPC barrier
is strongly spin dependent in all three cases, with the chemical
potential $\mu$ lying above the spin-up barrier but below the
spin-down barrier.  These observations are
consistent with the results of Wang and Berggren\cite{Wang98}.

In Fig.~1(d)-(f) we show the 1D electron density in the QPC.
The solid lines give the net spin-up density and the dashed lines
give the spin-averaged densities, which approach
$n_{1D}/2=1.40\times 10^{-2}{\rm nm}^{-1}$ far into the wire.
For all three QPC lengths, there is an excess spin-up density in
the vicinity of the barrier, with Friedel oscillations
persisting into the wire. In each case,
the integrated net spin-up density is close to one spin:
0.85 for (a), 0.93 for (b), and 0.90 for (c). Thus there is
a local moment with a spin of 1/2 formed at the QPC.

The local moment found within SDFT results from
the self-consistent flattening of the QPC barrier.
For example, above a square barrier there is a series
of quasi-bound states, resulting from multiple reflections
from the edges of the barrier.
While the bare QPC potential does not
give rise to resonances, the self-consistent potential
is flattened for spin-up, particularly for the longer QPCs, and
the first resonance in the spin-up local density of states is
clearly resolved. Since this resonance lies below the chemical
potential, it is fully occupied. The result is a localized
net spin of 1/2 at the QPC.

An important question is whether SDFT is reliable in predicting
a local moment. SDFT is a mean-field theory, and has limited
utility for strongly correlated electron systems.
In particular, consider an Anderson model consisting of
a single site with a Coulomb interaction $U$ between two
spin-degenerate orbitals of energy $\varepsilon_0$,
with hybridization $\Gamma$ to band electrons\cite{Anderson}.
For partially occupied orbitals, the
local density of states for each spin splits into two peaks
one at $\varepsilon_0$ and the other at $\varepsilon_0 +U$
(and at low temperatures an additional Kondo peak at the
chemical potential)\cite{Hewson}. Taken at face value,
SDFT is inadequate for this model as it can give only one
peak for each spin. However, SDFT has been applied successfully 
to calculate local moments in a system closely corresponding to 
the Anderson model, namely transition-metal adatoms on 
surfaces\cite{Stepanyuk}. The key is that even though the SDFT 
solution breaks spin-rotation symmetry, the frozen magnetization
can still be reliably interpreted as the {\it magnitude} of the 
dynamic local moment. In practice, we follow the
SDFT method developed by Janak to find the magnetization of 
metallic metals\cite{Janak}. Specifically, we first
solve the SDFT equations in a polarizing
magnetic field, and then reduce the field to zero
to extract the parameters $U$ and $\Gamma$ of the
underlying Anderson model.

To estimate the properties of the bound state, we go beyond
the formal validity of SDFT and study the Kohn-Sham orbitals.
Specifically, we plot the local 1D density of states
$\nu_{\sigma}(\epsilon)$ at the center of the QPC
in the insets of Fig.~1(a-c),
for the respective physical parameters.
Below the chemical potential,
the local density of states $\nu_{\uparrow}(\epsilon)$
for spin-up electrons shows a resonance
which broadens as the QPC is shortened.

Figure~2 depicts the local 1D density of states $\nu_{\sigma}(\epsilon)$
for $\hbar\omega_x=1.5 {\rm meV}$ in the presence of in-plane
magnetic fields $B$ from 0T to 10T in steps of 1T\cite{inplaneB}. Traces
are
vertically offset by a constant amount. The solid lines are for spin-up
and the dashed lines are for spin-down electrons.  With increasing
Zeeman splitting, the resonance for spin-up electrons shifts to
lower energies while the onset of the spin-down density of states
shifts to higher energies. From the residual splitting
between these features at $B=0$ we obtain an estimate for
the Coulomb-interaction energy $U$ in a local-moment description
of the QPC. Similarly, we obtain the hybridization $\Gamma$
from the width of the spin-up resonance.

Fig.~3 shows the dependence of the Coulomb energy $U$ and
the hybridization $\Gamma$ on the length of the QPC.
$U$ is obtained from the energy difference
between the resonance center of $\nu_{\uparrow}(\epsilon)$ and the
energy
at which the derivative of $\nu_{\downarrow}(\epsilon)$ is a maximum.
$\Gamma$ is obtained from the FWHM of a Lorentzian fit to the resonance
in $\nu_{\uparrow}(\epsilon)$.
The hybridization energy $\Gamma$ increases sharply
from $\Gamma\simeq 0.1{\rm meV}$ up to $\simeq 0.6{\rm meV}$ as the
QPC is shortened from $d = 80 {\rm nm}$ to $d = 40 {\rm nm}$.
The Coulomb energy $U$, on the other hand, stays nearly constant at
$\simeq 0.6{\rm meV}$ over this range.

Within the Anderson model, a strong Kondo peak in the differential
conductance is expected when $kT < k_B T_K < \Gamma \ll U$. This
suggests
that there is an optimal range of QPC lengths for observing the
Kondo effect: for very long QPCs, $\Gamma$ and therefore the Kondo
temperature $T_K$ will be too small compared to the real temperature,
whereas for very short QPCs, the resonance width $\Gamma$ will be
too broad for a Kondo effect ever to develop.
Experimentally, Reilly {\it et al.}\cite{Reilly} observed that the
extra conductance plateau in QPCs decreased from $0.7 (2e^2/h)$
to $0.5(2e^2/h)$ with increasing QPC length, suggesting a suppression
of the Kondo effect\cite{Meir}. Our SDFT results
suggest that it is primarily the decreasing hybridization of
the local moment to the leads that is responsible for the decrease
of the Kondo temperature and the evolution of the conductance plateau.

We have presented SDFT results only for chemical potentials
sufficiently above the resonance energy that the
local moment is fully formed. SDFT is generally unreliable for
a partially filled orbital, unless self-interaction effects
are explicitly removed\cite{selfint}, and this is not practical
for an open system with long range interactions like a QPC.

In conclusion, we studied the electronic states of quantum point
contacts (QPCs) using the spin-density-functional method. We found that
a
local moment with a spin of 1/2 is formed in the vicinity of the QPC
barrier.  This strongly supports recent claims of a Kondo effect
in transport through a quantum point contact\cite{Cronenwett}.
For the local moment, we obtained estimates for both the onsite
Coulomb-interaction energy and the hybridization to the leads.
The latter decreases rapidly with increasing length of the QPC.

Work by Y.M. at Rutgers was
partially supported by NSF grant DMR 00-93079 and by
DOE grant DE-FG02-00ER45790.

$^\dagger$ Permanent Address:
Department of Physics, Ben-Gurion University, Beer Sheva
84105, Israel.

\noindent
{\bf Figure Captions}
\begin{itemize}

\item[Fig.1:]
Results for QPC potentials of increasing sharpness: (a),(d)
$\hbar \omega_x = 1.0 {\rm meV}$, (b),(e)
$\hbar \omega_x = 1.5 {\rm meV}$, and (c),(f)
$\hbar \omega_x = 2.0 {\rm meV}$.
(a)-(c) Self-consistent ``barrier", {\it i.e.} energy of the bottom of
the lowest 1D subband $\epsilon_0(x)$ at temperature $T=0.1K$ as a
function of position $x$ in the direction of current flow through QPC.
The chemical potential $\mu$ is indicated by an arrow on the left. Solid
lines are for spin-up and dashed lines are for spin-down electrons.
Insets: local density of states
$\nu_{\sigma}(\epsilon)$ at the center of the QPC. (d)-(f) 1D electron
density in QPC. The solid line gives the net spin-up density and the
dashed line gives the spin-averaged density. Insets: contour plot of the
QPC potential $V_{\rm QPC}(x,y)$.
\item[Fig.2:]
Local density of states at the center of the QPC for
in-plane magnetic fields $B_{||}$ from 0 T to 10 T
in steps of 1 T\protect\onlinecite{inplaneB}.
Traces are vertically offset by a constant amount. The solid
lines are for spin-up and the dashed lines are for spin-down electrons.
\item[Fig.3:]
Hybridization energy $\Gamma$ and on-site Coulomb energy $U$ as a
function of
$\hbar\omega_x$, the sharpness of the external QPC potential.
$\Gamma$ is obtained from the FWHM of the Lorentzian fit to
the resonance. $U$ is obtained from the energy difference between the
resonance center and the maximum of $d\nu(\epsilon)/d\epsilon$
for the high-energy spin. Inset: $\nu_{\uparrow}(\epsilon)$ and
$d\nu_{\downarrow}(\epsilon)/d\epsilon$ at the center of QPC
for $\hbar \omega_x = 1.5 {\rm meV}$.
The dashed line is the Lorenzian fit to the resonance of
$\nu_{\uparrow}(\epsilon)$.
Hybridization energy $\Gamma$ and on-site Coulomb energy $U$ are
indicated.
\end{itemize}


\begin{references}

\bibitem{Thomas} K.~J.~Thomas {\it et al.}, Phys.~Rev.~Lett. {\bf 77},
135 (1996);
K.~J.~Thomas {\it et al.}, Phys.~Rev.~{\bf B58}, 4846 (1998).
\bibitem{Kristensen} A.~Kristensen {\it et al.}, Phys.~Rev. {\bf B62},
10950 (2000).
\bibitem{Nuttinck} S.Nuttinck {\it et al.}, Jpn.~J.~App.~Phys., {\bf
39}, L655 (2000). K.Hashimoto {\it et al.}, {\it ibid} {\bf 40}, 3000
(2001).
\bibitem{Reilly} D.~J.~Reilly {\it et al.}, Phys.Rev. {\bf B63}, 121311
(2001).
\bibitem{vanWees} B.~J.~van~Wees {\it et al.}, Phys.~Rev.~Lett. {\bf
60}, 848 (1988).
\bibitem{Wharam} D.~A.~Wharam {\it et al.}, J.~Phys.~C {\bf 21}, L209
(1988).
\bibitem{Wang98} C.-K.~Wang and K.-F.~Berggren, Phys.~Rev. {\bf B57},
4552 (1998).
\bibitem{Flambaum} V.~V.~Flambaum and M.~Y.~Kuchiev, Phys.~Rev. {\bf
B61}, R7869 (2000).
\bibitem{Spivak} B.~Spivak, and F.~Zhou, Phys.~Rev. {\bf B61}, 16730
(2000).
\bibitem{Rejec} T.~Rejec and A.~Ramsak, Phys.~Rev. {\bf B62}, 12985
(2000).
\bibitem{Bruus} H.~Bruus, V.~V.~Cheianov, and K.~Flensberg, Physica~E
{\bf 10}, 97 (2001).
\bibitem{Tokura} Y.~Tokura and A.~Khaetskii, Physica~E {\bf 12}, 711
(2002).
\bibitem{Cronenwett} S.~M.~Cronenwett {\it et al.}, Phys.~Rev.~Lett.
{\bf 88}, 226805 (2002).
\bibitem{inplaneB} A magnetic field in the plane of the QPC
induces a Zeeman splitting between spin-up and spin-down electrons
without affecting their orbital degrees of freedom.
\bibitem{Meir} Y.~Meir, K.~Hirose and N.~S.~Wingreen, Phys.~Rev.~Lett. {\bf
89}, 196802 (2002).
\bibitem{Lieb} E.~Lieb and D.~Mattis, Phys.~Rev. {\bf 125}, 164 (1962).
\bibitem{Callaway} J.~Callaway and N.~H.~March, Solid State Phys. {\bf
38}, 135 (1984).
\bibitem{Hirose} K.~Hirose, S.~S.~Li, and N.~S.~Wingreen,
Phys.~Rev. {\bf B63}, 033315 (2001); K.~Hirose and
N.~S.~Wingreen, {\it ibid}.~{\bf B64}, 073305 (2001).
\bibitem{Kohn} W.~Kohn and L.~J.~Sham, Phys.~Rev. {\bf 140}, A1133
(1965).
\bibitem{local} We use the parameterized form by Tanatar and Ceperley
for the two-dimensional electron gas for the local exchange-correlation
energy.
[B.~Tanatar and D.~M.~Ceperley, Phys.~Rev. {\bf B39}, 5005 (1989)].
\bibitem{Hirose2} K.~Hirose and M.~Tsukada, Phys.~Rev. {\bf B51}, 5278
(1995).
\bibitem{Anderson} P.~W.~Anderson, Phys.~Rev. {\bf 124}, 41 (1961).
\bibitem{Hewson} A.~C.~Hewson, {\it The Kondo Problem to Heavy Fermions}
(Cambridge University Press, Cambridge, 1997).
\bibitem{Stepanyuk} V.~S.~Stepanyuk {\it et al.}, Phys.~Rev.
{\bf B53}, 2121 (1996).
\bibitem{Janak} J.~F.~Janak, Phys.~Rev. {\bf B16}, 255 (1977).
\bibitem{selfint} J.~P.~Perdew and A.~Zunger, Phys.~Rev. {\bf B23}, 5048
(1981).
\end{references}
\end{document}